\documentclass[conference]{IEEEtran}

\usepackage{cite}
\usepackage{amsmath,amssymb,amsfonts}
\usepackage{algorithmic}
\usepackage{graphicx}
\usepackage{textcomp}
\usepackage{xcolor}
\def\BibTeX{{\rm B\kern-.05em{\sc i\kern-.025em b}\kern-.08em
    T\kern-.1667em\lower.7ex\hbox{E}\kern-.125emX}}
\usepackage{subcaption}
\usepackage[bookmarks=false]{hyperref}
\usepackage[T1]{fontenc}
\newcommand{\mb}[1]{{\mbox{\emph{#1}}}}

\begin{document}

\title{Retinopathy of Prematurity Stage Diagnosis Using Object Segmentation and Convolutional Neural Networks \\
\author{\IEEEauthorblockN{Alexander Ding$^{1}$, Qilei Chen$^{2}$, Yu Cao$^{2}$, Benyuan Liu$^{2}$}
\IEEEauthorblockA{$^1$ Commonwealth School, Boston, MA, USA\\
$^2$ Department Of Computer Science, University of Massachusetts Lowell, Lowell, MA, USA\\
alding@commschool.org, qilei\_chen@student.uml.edu, \{ycao, bliu\}@cs.uml.edu}}
\thanks{}
}

\maketitle



\begin{abstract}
Retinopathy of Prematurity (ROP) is an eye disorder primarily affecting premature infants with lower weights. It causes proliferation of vessels in the retina and could result in vision loss and, eventually, retinal detachment, leading to blindness. While human experts can easily identify severe stages of ROP, the diagnosis of earlier stages, which are the most relevant to determining treatment choice, are much more affected by variability in subjective interpretations of human experts. In recent years, there has been a significant effort to automate the diagnosis using deep learning. This paper builds upon the success of previous models and develops a novel architecture, which combines object segmentation and convolutional neural networks (CNN) to construct an effective classifier of ROP stages 1-3 based on neonatal retinal images. Motivated by the fact that the formation and shape of a demarcation line in the retina is the distinguishing feature between earlier ROP stages, our proposed system first trains an object segmentation model to identify the demarcation line at a pixel level and adds the resulting mask as an additional "color" channel in the original image. Then, the system trains a CNN classifier based on the processed images to leverage information from both the original image and the mask, which helps direct the model's attention to the demarcation line. In a number of careful experiments comparing its performance to previous object segmentation systems and CNN-only systems trained on our dataset, our novel architecture significantly outperforms previous systems in accuracy, demonstrating the effectiveness of our proposed pipeline.
\end{abstract}


\section{Introduction}
\begin{figure*}[t]
    \centering
    \begin{subfigure}[t]{0.23\textwidth}
        \centering
        \includegraphics[width=4cm]{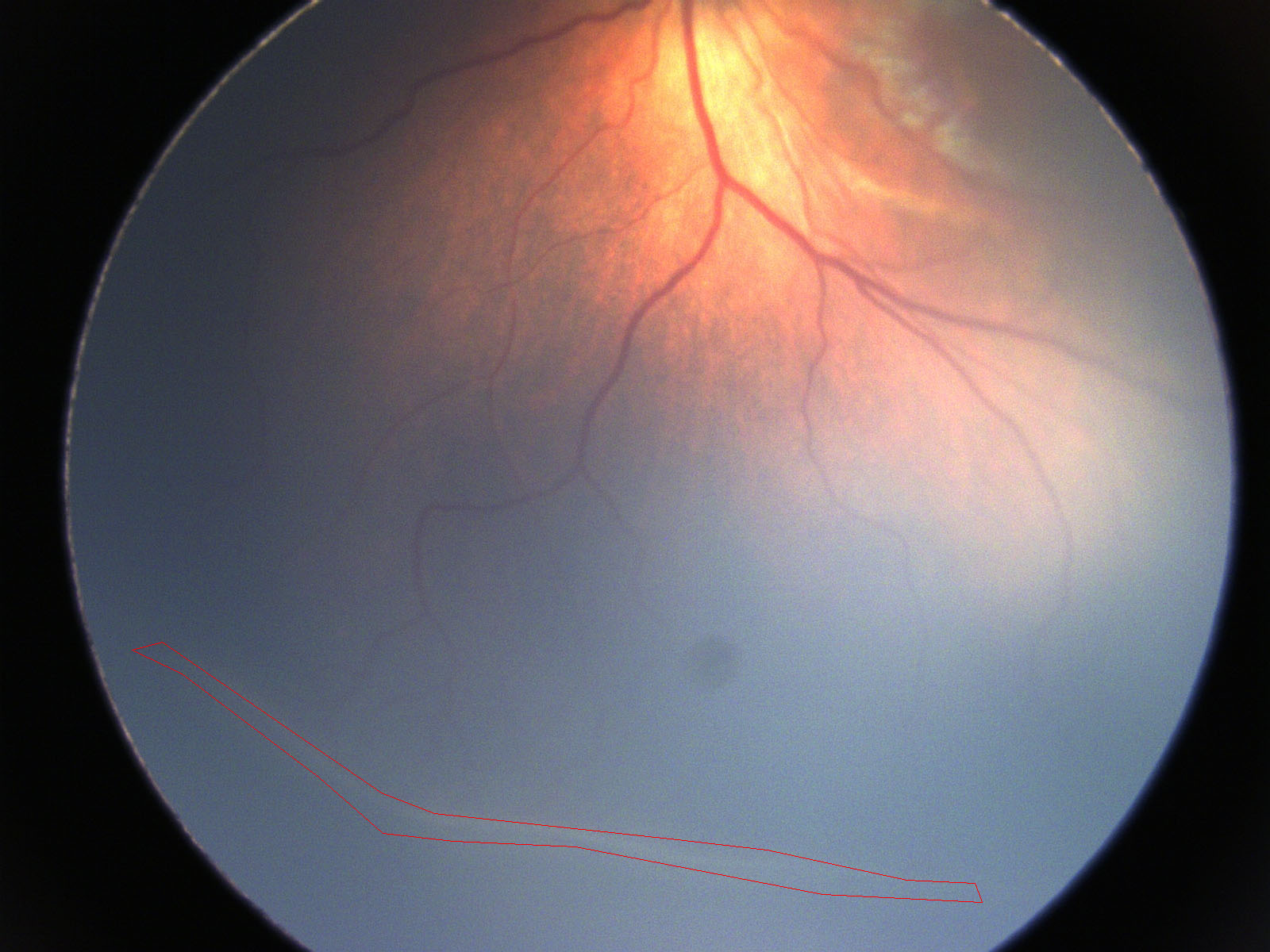}
        \caption{Stage 1}
    \end{subfigure}%
    ~
    \begin{subfigure}[t]{0.23\textwidth}
        \centering
        \includegraphics[width=4cm]{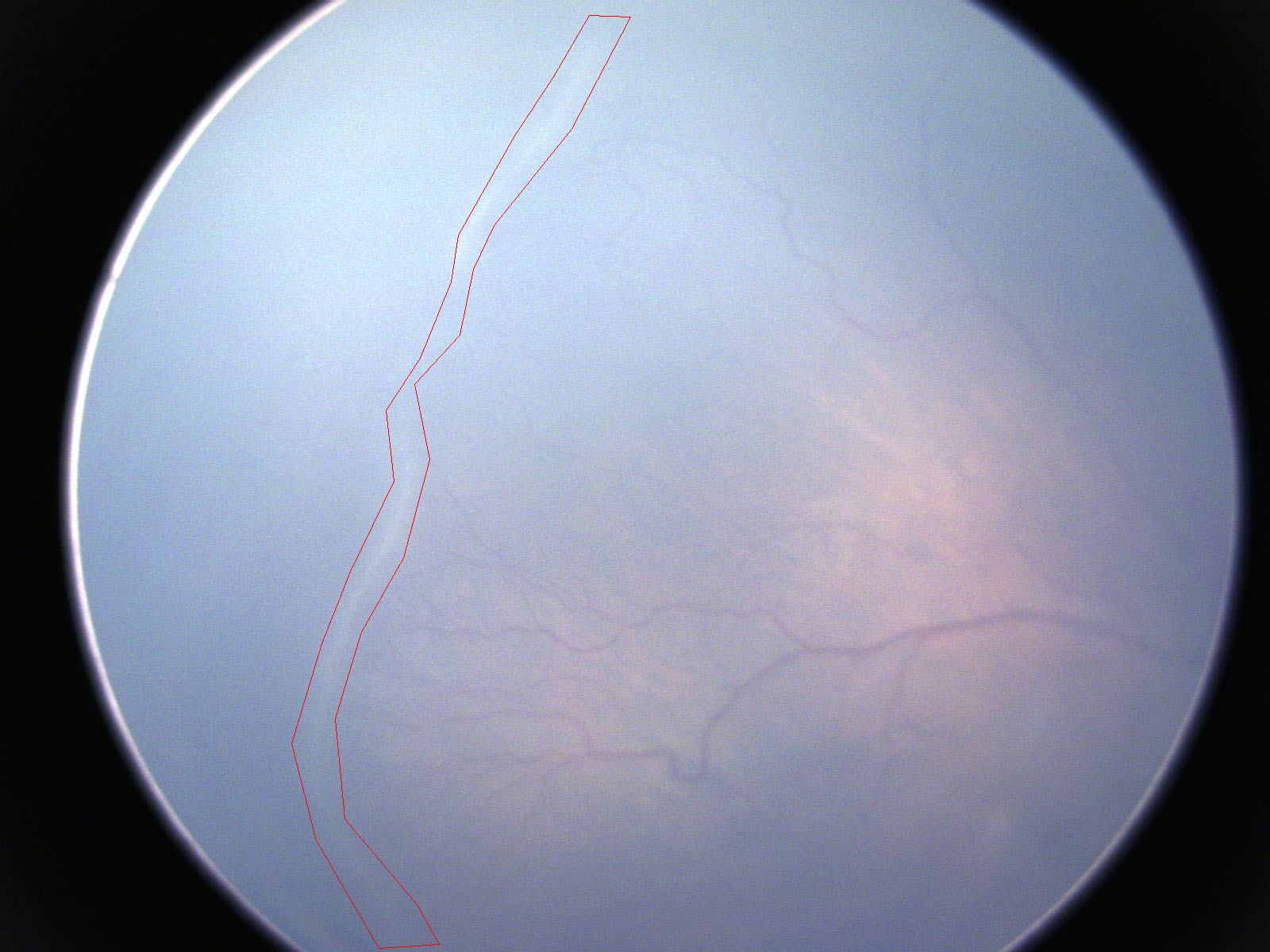}
        \caption{Stage 2}
    \end{subfigure}%
    ~
    \begin{subfigure}[t]{0.23\textwidth}
        \centering
        \includegraphics[width=4cm]{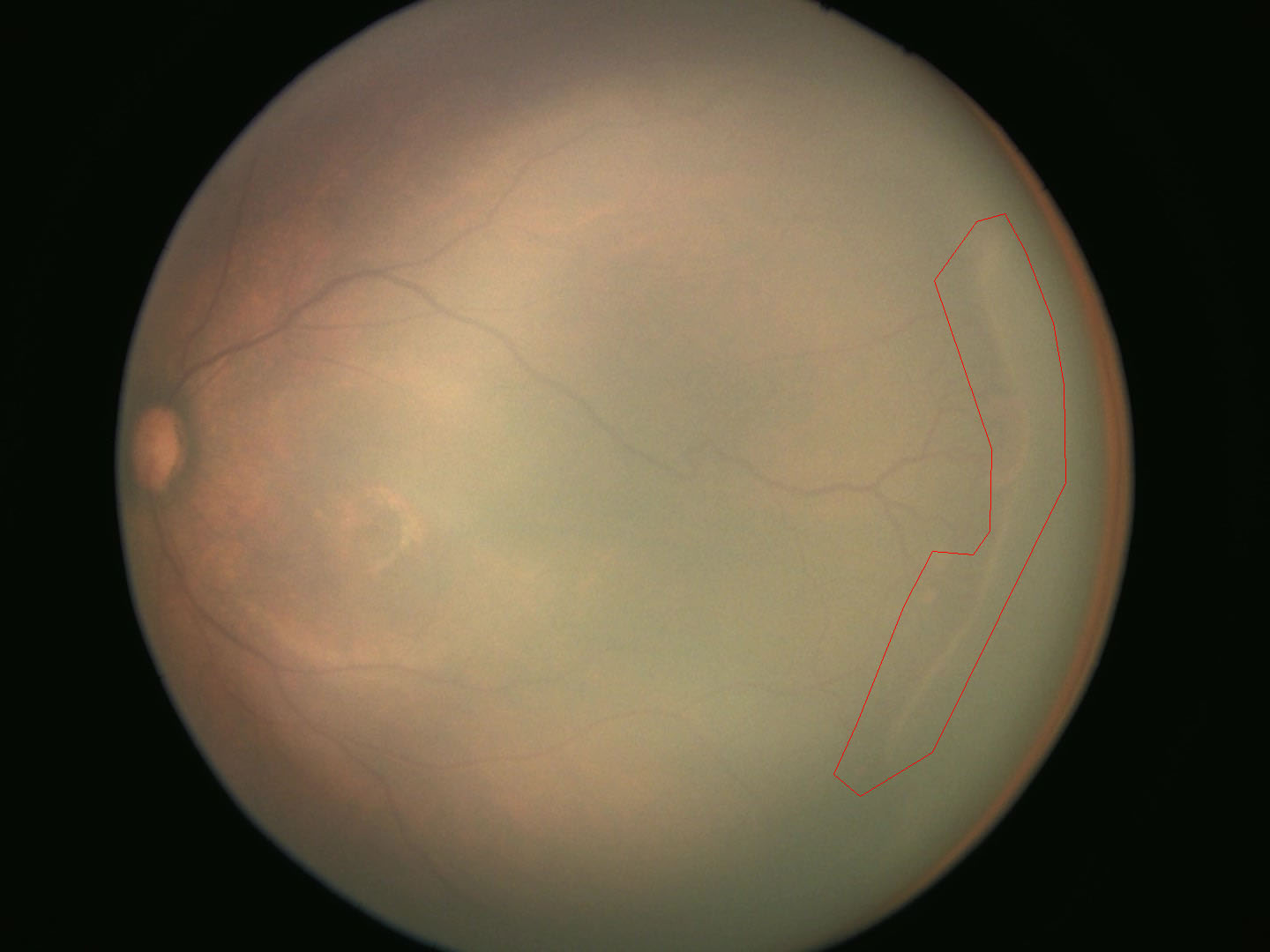}
        \caption{Stage 3}
    \end{subfigure}%
    ~
     \begin{subfigure}[t]{0.23\textwidth}
        \centering
        \includegraphics[width=4cm]{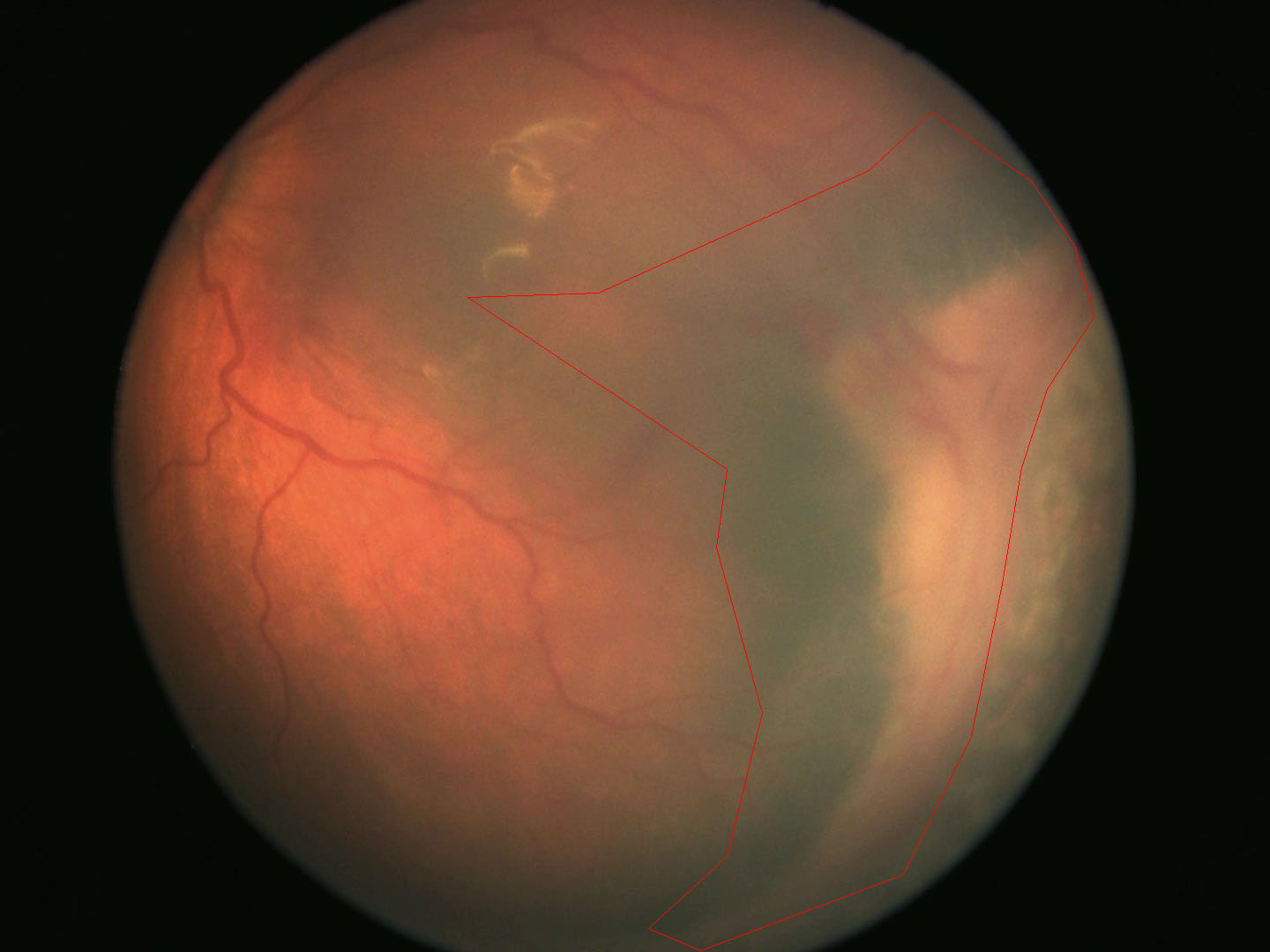}
        \caption{Stage 4}
    \end{subfigure}%
\caption{Retina images of the stages 1-4 of ROP, from left to right, respectively. Stage 5 indicates total retina detachment and therefore does not have a retina image. The red marker represents the annotated polygons bounding the demarcation line, which looks like a thin white line. (a) Stage 1: demarcation line between vascular and avascular areas of the retina. (b) Stage 2: intraretinal ridge (which is simply the demarcation line elevated). (c) Stage 3: ridge with ragged posterior due to abnormal vessel growth outside of the retina toward the center of the eye. (d) Stage 4: partial retina detachment. 
\label{fig:rop-stages}}
\end{figure*}

Retinopathy of Prematurity (ROP) is a retina disorder that is observed primarily in premature infants with lower weights. ROP results in abnormal blood vessel growth in the retina, potentially leading to scarring and retina detachment. ROP ranges from mild, where the disease often resolves on its own and results in normal vision development, to severe, where potential retina detachment could blind the infant should further treatment not be taken. The diagnosis of ROP is divided into 5 stages, described in Figure \ref{fig:rop-stages}.

ROP can also be classified based on the characteristics of the blood vessels in the posterior pole as normal, plus, pre-plus, or aggressive posterior ROP (AP-ROP). Plus ROP is recognized by the increased venous dilatation and arteriolar tortuosity of the vessels, while pre-plus ROP is in between normal and plus, and AP-ROP is a rapidly progressing form of plus ROP \cite{eyewiki}. There has been significant work to automate the process of diagnosing these labels using deep learning methods, such as the work of \textit{Tan et al.} and \textit{Brown et al.} \cite{tan2019deep} \cite{brown2018automated}. However, this paper is only concerned with the diagnosis of ROP stages and therefore does not give them further discussions. 

The diagnosis of ROP, for both human experts and deep learning programs, entails inspecting high resolution images of the patient retinas captured by imaging systems like RetCam3. However, human diagnosis is limited in several respects. First, the qualitative nature of the stage definitions means that diagnosis relies mostly on the subjective interpretations of ophthalmologists. Naturally, this has resulted in high inter-expert and intra-expert variability in clinical classifications, especially for Stages 1-3 \cite{gschlieber}. Second, many developing nations with high population density, such as China and India, lack enough trained ophthalmologists to match the huge number of premature infants. Moreover, there are currently 14 countries in the world that report no ROP screening procedure \cite{adams_2019}. The high barrier of entry to ROP screening has left many cases of preventable blindness untreated. 

This paper focuses on the diagnosis of Stages 1-3 ROP for several reasons. First, whereas Stage 4 and 5 are immediately identifiable given the severity of the physical symptoms, Stages 1-3 and normal retinas are more subtly classified by the existence, size, and shape of the demarcation line (or ridge, in later stages---we shall simply refer to them demarcation line for conciseness), as well as vascular proliferation. Moreover, it is more pressing to diagnose between Stages 1-3, since patients suffering from Stages 4-5 ROP have already sustained irreversible damage to the retina, while the diagnosis between Stages 1-3 of ROP is critical in allowing doctors to recommend the appropriate treatment while blindness is still preventable. 

This paper proposes an improved, hybrid architecture over the naive application of CNNs and object segmentation models to classify Stages 1-3 of ROP. Instead of simply feeding the original images into a CNN, we first run an object segmentation model on the input images to localize the demarcation lines, generating a binary mask. Then, we combine the mask with the original image (which is preprocessed into a one channel black and white image) as an additional channel to feed a 2-channel input into a traditional CNN architecture for classification. By adding an independent object segmentation model to highlight areas of interest for our CNN, we are providing additional information that a CNN is not well-structured to generate on its own. 

To demonstrate our hybrid architecture's effectiveness, we run a number of experiments. Besides testing the architecture itself, we also separately train and evaluate its individual components---a CNN-only system and an object segmentation system---using the same setups and compare the results. Our hybrid architecture achieves an increase in overall accuracy by 13\% and 20\% from the CNN-only system and the object segmentation system, respectively. This result shows that the additional information about demarcation lines focuses the classifier's attention on distinguishing features, and the combination of a CNN and an object segmentation system is able to make more robust predictions than its individual components. 

The rest of the paper is structured as follows. Section \ref{sec:related-work} describes related work on using deep learning methods such as CNN, object segmentation, and fully convolutional network to classify ROP stages. Section \ref{sec:dataset-and-method} is dedicated to the the dataset used in our experiments and our methodologies, including image preprocessing, the pipeline of our architecture, its component models, and our use of transfer learning. Section \ref{sec:experiment} documents a number of experiments on our hybrid model, details the training setups, and discusses the results. Finally, section \ref{sec:conclusion} concludes this paper by summarizing the contribution of our work and proposing several directions for future work. 

\section{Related Work}
\label{sec:related-work}

There has been a number of proposed automated systems to assist ophthalmologists in their diagnosis of ROP stages. \textit{Hu et al.} proposes the usage of a CNN to classify ROP stages \cite{hu_et_al}. The determination of the stage of ROP, however, heavily depends on the demarcation line, which constitutes an extremely small percentage of the overall image. CNNs are well-suited for classifications based on large parts of an image but often struggle to focus exclusively on small details. 

Following a separate approach, \textit{Mulay et al.} utilizes the Mask R-CNN architecture to make pixel-level segmentation of the demarcation lines in retina images and used the segmentation result (presence of demarcation lines) to supply a binary classification between normal and ROP, though it does not classify the exact stage \cite{mulay_ram_sivaprakasam_vinekar_2019}. 

In a recent paper, \textit{Chen et al.} proposes a hybrid architecture, which uses a fully convolutional network (FCN) to generate a pixel-level binary segmentation map, which is fed into a multi-instance learning (MIL) module along with the original image to classify the stage of ROP \cite{chen_zhao_zhang_wang_zhang_lei_2019}. The paper conducts experiments comparing the performance of the MIL module alone and the performance of the hybrid FCN+MIL setup. By recognizing the importance of helping the classifier focus on distinguishing feature, as well as utilizing multiple images for each patient, the hybrid setup achieves $92.25\%$ accuracy for classification between Stages 1-4 and normal retina images, a $5.36\%$ increase in accuracy compared to the MIL module alone. Of the related works discussed, the overall idea of this architecture is most similar to the one presented in this paper. Though much of our work is conducted before this paper is published, our object segmentation masking can be used in conjunction with the FCN segmentation map to further improve the performance of the MIL module. 
\section{Dataset and Method}
\label{sec:dataset-and-method}
\subsection{Data}
\label{sec:data}

We obtained a dataset of retina images of ROP patients captured by the RetCam3 imaging system in collaboration with a hospital. The original images have dimensions (in pixels) $1600\times1200$, $1440\times1080$, $640\times480$, and $720\times480$, though they are resized to $299\times299$ when inputting to our deep learning models. 

Each image is annotated by one experienced doctor using the Visual Geometry Group (VGG) Image Annotator \cite{dutta2019vgg}. The doctor draws a bounding polygon surrounding the demarcation line on the retina and labels the polygon with the stage it represents. That constitutes our object segmentation model's dataset. We synthesize a classifier dataset from the same source images by dividing them into their respective stages, identified by the polygon's label in each image. Both datasets are partitioned into train, test, and validation subsets in a $6:3:1$ split. The exact split of the data is given by Table \ref{tab:datasets}.

\begin{figure}[ht]
    \centering
    \begin{subfigure}[t]{0.23\textwidth}
        \centering
        \includegraphics[width=4cm]{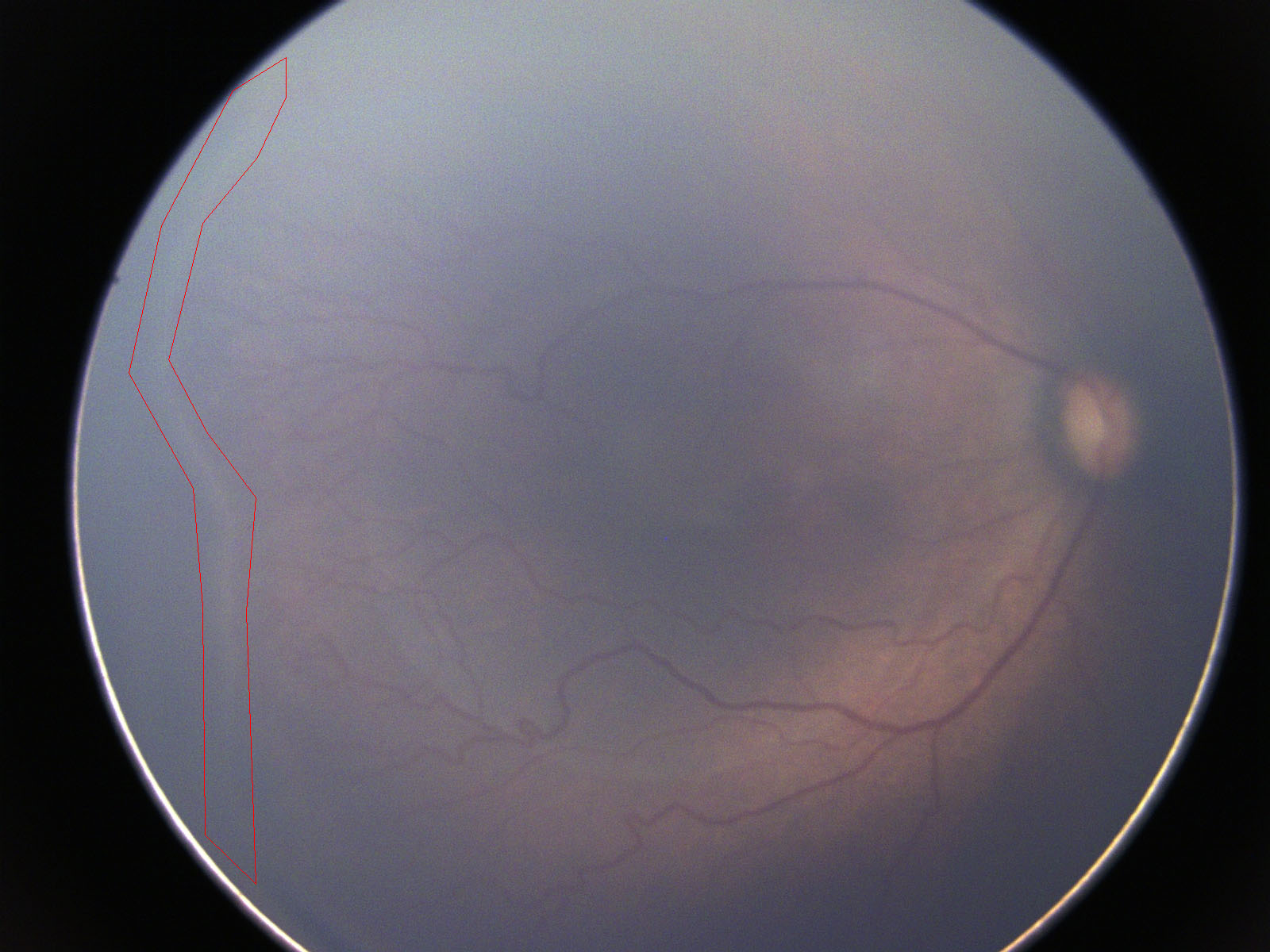}
    \end{subfigure}%
    ~
    \centering
    \begin{subfigure}[t]{0.23\textwidth}
        \centering
        \includegraphics[width=4cm]{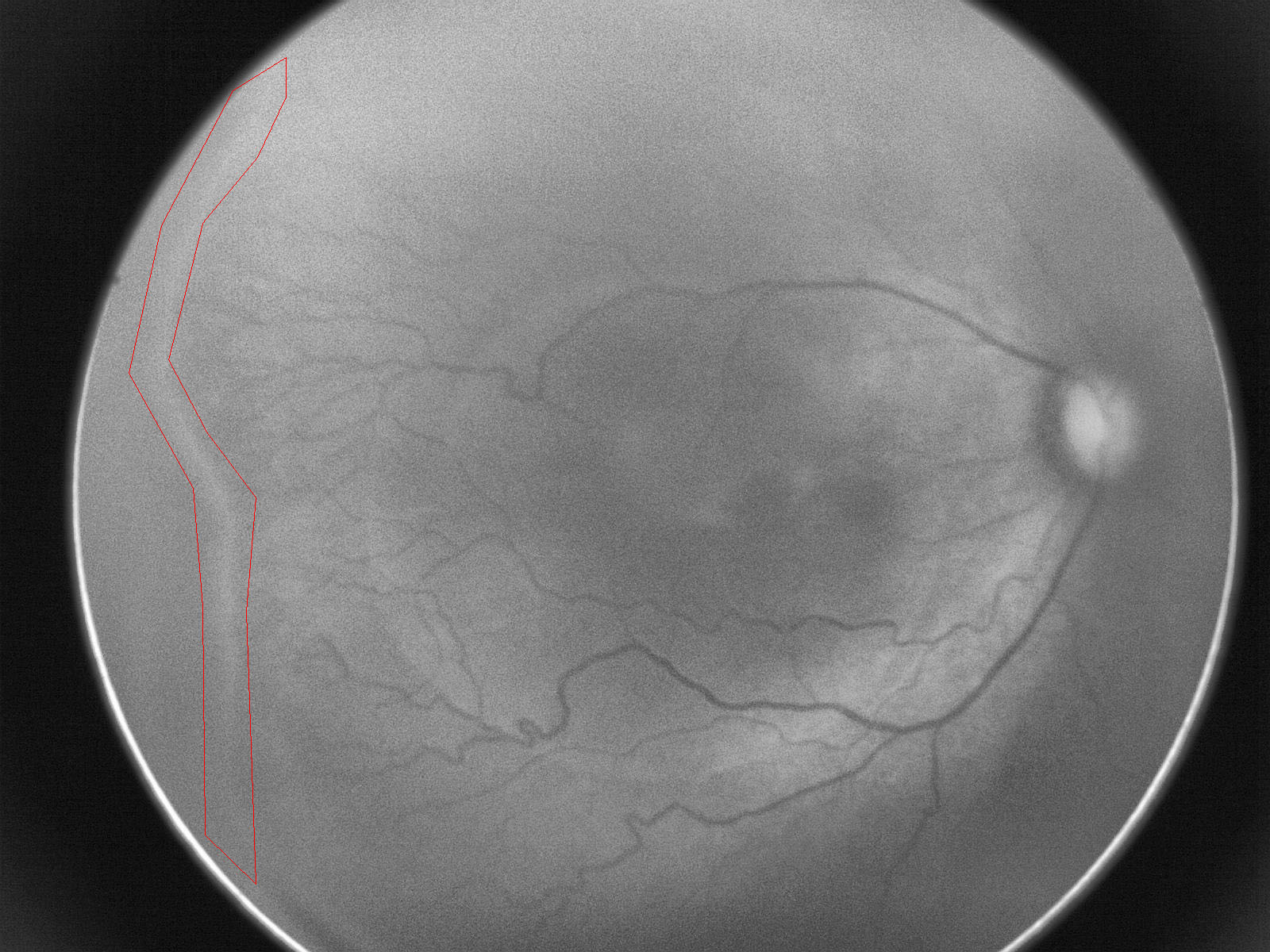}
    \end{subfigure}%

\caption{Example of the effects of preprocessing. The original image (left) is blurry and difficult to read, while the preprocessed image (right) accentuates line features, in particular the demarcation line. }
\label{fig:preprocessing}
\end{figure}

\begin{table}[ht]
    \centering
    \caption{ROP Datasets}
    \begin{tabular}{|c|c|c|c|c|c|c|}
        \hline
        & \multicolumn{3}{c|}{\textbf{Without Augmentation}} & \multicolumn{3}{c|}{\textbf{With Augmentation}} \\
        \cline{2-4} \cline{5-7}
                 & \textbf{\textit{Train}} & \textbf{\textit{Test}} & \textbf{\textit{Validation}} & \textbf{\textit{Train}} & \textbf{\textit{Test}} & \textbf{\textit{Validation}} \\
        \hline
        Stage 1  & 141 & 70  & 23 & 705 & 70  & 23 \\
        Stage 2  & 796 & 398 & 132 & 796 & 398 & 132 \\
        Stage 3  & 720 & 360 & 119 & 720 & 360 & 119 \\
        \hline
    \end{tabular}
    \label{tab:datasets}
\end{table}

Note that the original dataset (without augmentation) suffers from an imbalance between different classes. In order to mitigate this bias and give equal attention to each class, we augment Stage 1 images by a factor of 5. Specifically, we duplicate each Stage 1 image 4 times. For each copy, we randomly crop between $[0\%, 5\%]$ of the image along each edge and zoom into the center of the cropped image by a random factor between $[100\%, 110\%]$. This method yields slightly different copies of the same image to balance the classes. Variations of the same image are kept in the same train, test, or validation dataset so as to prevent the models overfitted to the training dataset from scoring an unrepresentatively high accuracy during validation or testing.  

The original images are plagued with low-contrast lighting, blurry line features, and stylistic differences between images from different sources. The blurry line features impair both human and machine interpretation of the images \cite{sevik_kose_berber_erdol_2014}. Before using the images for training, we preprocess the images to standardize input and accentuate line features. We first grayscale the image to remove any coloring distinctions unique to imaging equipment that may distract the model's learning, given by
\begin{equation}
    I' \equiv ((0.3 \times R) + (0.59 \times G) + (0.11 \times B) ),
\end{equation}
where R, G, B are the red, green, and blue channels, respectively and I' is the new image. 

Then, we apply histogram equalization to the grayscale image to increase its contrast, implemented in OpenCV \cite{histogram_equalization_2019}. Histogram equalization creates a histogram of the image's pixels based on intensity, defined by $H(i)$ for intensity $i$, $i \in [0,255]$. Then, it transforms the histogram into a new histogram, given by
\begin{equation}
    H'(i) \equiv \Sigma_{0 \leq j < i}{H(j)}.
\end{equation}
It then scales the new histogram $H'$ back into the $[0,255]$ range and remaps the pixels to this new distribution by
\begin{equation}
    I'(x,y) \equiv H'(I(x,y)),
\end{equation}
where $I(x,y)$ refers to the image intensity of the pixel at coordinate $(x,y)$. 

In effect, histogram equalization increases the overall contrast of the image. After this step, we apply CLAHE (Contrast Limited Adaptive Histogram Equalization) to the resulting image, which divides the image into small tiles and applies histogram equalization to each tile \cite{zuiderveld1994graphics}. This method improves the global contrast of the image, but it would also amplify noises in tiles containing them. To avoid this, any histogram bin whose contrast is greater than a certain threshold is removed and spread out evenly among other bins before histogram equalization is applied. CLAHE finally applies bilinear interpolation to smooth out the tile borders.

After these steps of preprocessing unique to our application, we normalize the entire dataset by subtracting from each image its mean pixel value and dividing the result by the image's standard deviation, such that all values are in the range $[-1, 1]$. Figure \ref{fig:preprocessing} shows an example of the a preprocessed image side by side with its original counterpart. 

\subsection{Methodology}
\label{sec:methodology}

\subsubsection{Pipeline}
\label{sec:pipeline}

\begin{figure*}[t]
    \centering
    \includegraphics[width=\textwidth]{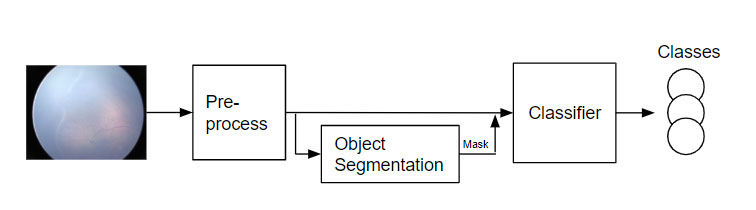}
    \caption{Pipeline of the hybrid architecture.}
\label{fig:pipeline}
\end{figure*}

The architecture of our pipeline is given by Figure \ref{fig:pipeline}. After preprocessing the images for both models, we train an object  segmentation model using the preprocessed images and their demarcation line annotations. We use the trained model to process a separate copy of the preprocessed dataset. For each image, the model generates a binary mask of the same shape highlighting the predicted demarcation line. We stack the mask with the preprocessed image into a two-channel image. This new dataset processed by the object segmentation model is used to train our classifier, which outputs our predictions. During inference, the image is preprocessed, fed to the object segmentation model to generate a mask, overlayed with the mask, and classified. 

\subsubsection{Object Segmentation Model}
\label{sec:oo}

We use the Mask R-CNN architecture for highlighting the demarcation lines at the pixel level \cite{1703.06870}. Mask R-CNN is an intuitive extension of Faster R-CNN \cite{girshick_2015}, which generates bounding boxes of detected objects in the image and classifies each bounding box. In addition to the region proposal network (RPN) and the classifier branches inherited from Faster R-CNN, Mask R-CNN adds a branch for generating binary pixel-level segmentation masks on each Region of Interest (ROI) by using a quantization-free \textit{RoIAlign} layer that faithfully translates between the compressed RoIs and their original coordinates in the image. These changes result in a much more robust and flexible architecture capable of not just object detection, but object segmentation. 

In this study, we use an open-source implementation of Mask R-CNN with Resnet-101-FPN backbone pretrained on the COCO dataset \cite{matterport_2019}. We are able to only modify the data input pipeline and hyperparameters and use the rest of the model as-is. Since our object segmentation model's purpose is to generate a binary mask for the original image regardless of the classification between demarcation lines of different stages of ROP, our Mask R-CNN model treats all demarcation lines as the same class of objects. 

In training, we employ transfer learning techniques to repurpose the pretrained model for our own classes. This reduces overfitting when training deep architectures on our small dataset, while still achieving good performances. Since many of the steps in image classification are common to all models regardless of the number of classes (like recognizing shapes, patterns, lines, etc.), transfer learning allows us to piggyback off of the pretrained weights' general image analysis abilities and fine-tune them for our specific use-case. When training, we unfreeze only the heads of each component, i.e., the RPN, the classifier, and the mask branch. 

\subsubsection{Classifier}
\label{sec:classifier}

For the classifier, we use a Inception v3 model (with average pooling) pretrained on ImageNet, provided by Tensorflow \cite{szegedy2016rethinking} \cite{imagenet_cvpr09}. To repurpose the pretrained model, which takes in shape $299\times299\times3$ RGB images and outputs 2048 classes of predictions, for our 2-channel images and 3 output classes, we modify the shape of the pretrained model's input layer and final dense layer and initialize them with random weights. Like the object segmentation model, we employ transfer learning to prevent overfitting and simplify the training process. When training, we unfreeze all layers. 

\begin{figure}[t]
    \centering
    \begin{subfigure}[t]{0.23\textwidth}
        \centering
        \includegraphics[width=4cm]{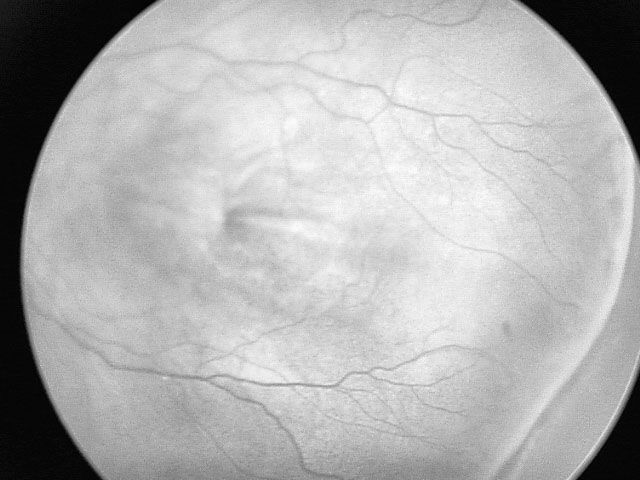}
    \end{subfigure}%
    ~
    \centering
    \begin{subfigure}[t]{0.23\textwidth}
        \centering
        \includegraphics[width=4cm]{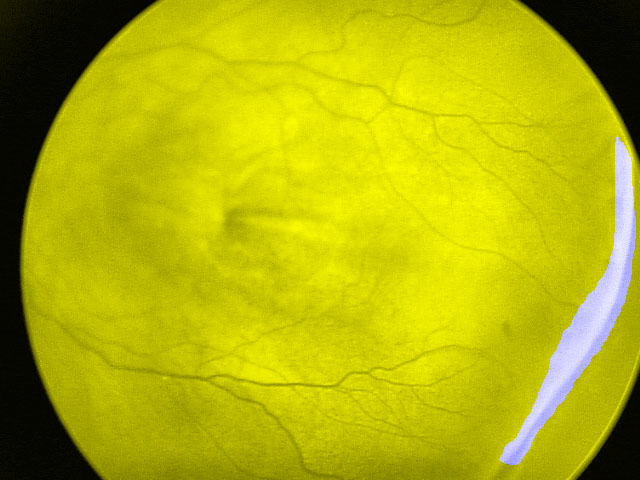}
    \end{subfigure}%
\caption{Example of combining the image and its generated binary mask. The original image (left) is simply the grayscale preprocessed image, containing one color channel, while the combined image (right) has the segmentation mask as one color channel (blue, in this case) and the original image as another (though for visualization purposes it constitutes both the red and the green channels.)}
\label{fig:overlay}
\end{figure}

\section{Experiment}
\label{sec:experiment}
\begin{table*}[t]
    \centering
    \caption{Test performance of the various architectures.}
    \label{tab:test-results}
    \begin{tabular}{|c|c|c|c|c|c|c|c|c|c|}
        \hline
        \textbf{Architecture} & \multicolumn{3}{c|}{\textbf{Hybrid}} & \multicolumn{3}{|c|}{\textbf{Classifier-Only}} & \multicolumn{3}{|c|}{\textbf{Object Segmentation}} \\
        \cline{2-4} \cline{5-7} \cline{8-10}
                & \textbf{\textit{Precision}} & \textbf{\textit{Recall}} & \textbf{\textit{F1-Score}} & \textbf{\textit{Precision}} & \textbf{\textit{Recall}} & \textbf{\textit{F1-Score}} & \textbf{\textit{Precision}} & \textbf{\textit{Recall}} & \textbf{\textit{F1-Score}} \\ 
        \hline
        Stage 1 & 0.78      & 0.77   & 0.78  & 0.98  & 0.36  & 0.53 & 0.60  & 0.05  & 0.09 \\
        Stage 2 & 0.61      & 0.62   & 0.61  & 0.45  & 0.86  & 0.59 & 0.44  & 0.60  & 0.51 \\
        Stage 3 & 0.62      & 0.62   & 0.62  & 0.59  & 0.36  & 0.45 & 0.57  & 0.73  & 0.64 \\
        \hline
        Accuracy & \multicolumn{3}{c|}{0.67} & \multicolumn{3}{c|}{0.54} & \multicolumn{3}{c|}{0.47} \\
        \hline
    \end{tabular}%
\end{table*}

%
%
%
%
%
%
%
%
%
%
\begin{table}[ht]
    \centering
    \caption{Confusion matrix of the hybrid architecture. Each row refers to the distribution of predictions for images of a particular stage; each column refers to the true stages of images predicted as a particular stage.}
    \label{tab:confusion-matrix-hybrid}
    \begin{tabular}{|c|c|c|c|}
        \hline
                & \textbf{Stage 1} & \textbf{Stage 2} & \textbf{Stage 3} \\
        \hline
        \textbf{Stage 1} & 271       & 56     & 23  \\
        \hline
        \textbf{Stage 2} & 39        & 245    & 114  \\
        \hline
        \textbf{Stage 3} & 37        & 98     & 225  \\
        \hline
    \end{tabular}%
\end{table}
\begin{table}[ht]
    \centering
    \caption{Confusion matrix of the classifier-only architecture. Refer to Table \ref{tab:confusion-matrix-hybrid} for how to read the results.}
    \label{tab:confusion-matrix-classifier}
    \begin{tabular}{|c|c|c|c|}
        \hline
                & \textbf{Stage 1} & \textbf{Stage 2} & \textbf{Stage 3} \\
        \hline
        \textbf{Stage 1} & 127       & 186    & 37  \\
        \hline
        \textbf{Stage 2} & 1         & 342    & 55  \\
        \hline
        \textbf{Stage 3} & 1         & 229     & 130  \\
        \hline
    \end{tabular}%
\end{table}
\begin{table}[ht]
    \centering
    \caption{Confusion matrix of the object segmentation model. The model predicts ROP-free when no demarcation line is identified. A ROP-free row is omitted as there is no ROP-free image (therefore, all ROP-free predictions are incorrect). Refer to Table \ref{tab:confusion-matrix-hybrid} for how to read the results.}
    \label{tab:confusion-matrix-oo}
    \begin{tabular}{|c|c|c|c|c|}
        \hline
                & \textbf{ROP-free} & \textbf{Stage 1} & \textbf{Stage 2} & \textbf{Stage 3} \\
        \hline
        \textbf{Stage 1} & 31 & 18        & 234    & 67  \\
        \hline
        \textbf{Stage 2} & 20 & 8         & 240    & 130  \\
        \hline
        \textbf{Stage 3} & 26 & 4         & 68     & 262  \\
        \hline
    \end{tabular}%
\end{table}
For all of the experiments, we use the same hyperparameters for each unique model. The object segmentation model---pretrained on COCO dataset---is retrained with learning rate of $0.002$, momentum of $0.9$, mini-batches of size $1$, and over $100$ epochs, saving the best weights based on validation loss. For inference, the confidence threshold is set to be $0.8$. The classifier is trained with learning rate of $0.005$, batch size of $32$, and an Adam optimizer. To further increase the effective size of our dataset without overfitting, we employ early stopping with a patience of $10$ monitoring the validation loss, saving the best weight. 

All the experiments are conducted on a Windows 10 machine, with a 4.00GHz Intel i7-6700K CPU, 16 GB of memory, and an NVIDIA GeForce GTX 1080 GPU, running Python 3.6 and Tensorflow v1.14.0 (with GPU support) and CUDA v10.0. 

In order to evaluate the effectiveness of our hybrid architecture, we compare its performance to its component models' performance in stage classification. Besides training the hybrid architecture itself, we train a separate classifier using the same setup as described in Section \ref{sec:classifier} without the binary mask overlay. In other words, this separate classifier conducts whole-image classification on the same preprocessed dataset without the addition of our object segmentation model. Likewise, we train a separate object segmentation model using the same setup as described in Section \ref{sec:oo}, except that demarcation lines of each stage are considered to be separate classes. In order to extract a stage classification out of an object segmentation model, we make a number of changes. For each image, the model generates potentially multiple object predictions, each with a predicted stage and a binary mask (i.e., where the demarcation line is); we union all predictions of the same stage into one binary mask and take the stage with the largest area to be the model's stage prediction. If no demarcation line is predicted, the model predicts the retina to be ROP-free, which is always incorrect in our tests. 

The results of our hybrid system, the classifier on its own, and the object segmentation model on its own are shown in Table \ref{tab:test-results}. For any stage, precision is computed as the rate of correct predictions in all predictions of images as that stage, while recall is the rate of correct predictions in all images of that stage. F1-score is subsequently defined as the harmonic mean of precision and recall, given by 
\begin{equation}
    F_1 \equiv 2 \cdot \frac{\mb{precision} \cdot \mb{recall}}{\mb{precision} + \mb{recall}}
\end{equation}
Finally, accuracy is simply the rate of correct predictions in all predictions. 

We also give the confusion matrices for the various architectures in Table \ref{tab:confusion-matrix-hybrid}, Table \ref{tab:confusion-matrix-classifier}, and Table \ref{tab:confusion-matrix-oo}. Each row refers to the distribution of predictions for images of that particular stage; as a result, each column refers to the true stages of images predicted as that particular stage. 

\begin{figure}[ht]
    \centering
    \includegraphics[width=0.46\textwidth]{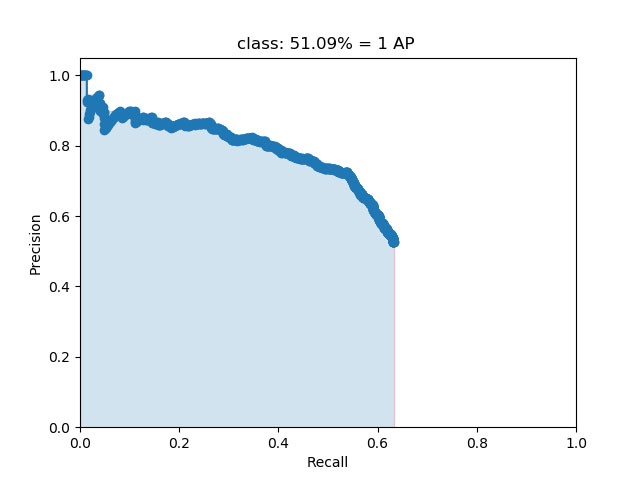}
    \caption{The precision-recall curve of the hybrid architecture's single class object segmentation model.}
\label{fig:precision-recall-curve}
\end{figure}

In addition, within our hybrid system, the single class object segmentation model achieves an average precision (AP) of $51.09\%$. Here, AP is defined as the area-under-curve of the precision-recall curve of the model, given in Figure \ref{fig:precision-recall-curve}. To compute the curve, we gather all predictions of demarcation lines, rank them from highest confidence level to lowest, and consider them one by one, keeping track of an aggregate precision and an aggregate recall value as we go. A prediction is considered correct if the intersection over union (IoU) value (defined as the intersection between the predicted area and the ground truth area over the union of the two) is greater than 50\%. Note that the recall value is strictly increasing in the process, which derives the curve. The curve ends before reaching 1.0 recall as the model fails to identify all of the demarcation lines. We compute mAP using an open-source script available on GitHub \cite{joan_mAP_2019}. 

Our hybrid system achieves a balanced precision/recall tradeoff and $67\%$ overall accuracy. It significantly outperforms the classifier-only setup, which only yields $54\%$ overall accuracy. Similarly, it outperforms the object segmentation system, as expected, since its architecture's primary purpose is to localize objects, not classify. Evidently, the demarcation line mask supplements our CNN with information it cannot otherwise compute, as the steps used to generate the resulting mask---such as region proposal and pixel level translation between ROI and original image---cannot be replicated or simulated using a pure CNN architecture. Therefore, by providing highlight to the demarcation lines, we are able to achieve a $13\%$ increase in accuracy. 

Moreover, the hybrid architecture's confusion matrix (Table \ref{tab:confusion-matrix-hybrid}) presents a much more even distribution of predictions than that of the classifier-only architecture (Table \ref{tab:confusion-matrix-classifier}), which is overfitted to Stage 2. The hybrid architecture is able to effectively classify between Stages 1 and 2, especially when compared to the classifier-only architecture. Clearly, the additional demarcation line highlighting has allowed the hybrid architecture to carefully distinguish between a demarcation line (for Stage 1) and a ridge (for Stage 2). The hybrid architecture is less robust between Stages 2 and 3, as both stages are distinguished only by slightly different ridges. As a result, the architecture achieves about $16\%$ higher precision and recall for Stage 1 than the other two stages. 

The object segmentation model struggles with detecting Stage 1 images, due to the low number of unaugmented Stage 1 training data. Augmentation does not benefit object segmentation models in any way, as cropping the images does not vary demarcation lines. Within Stage 2 and Stage 3 data, our hybrid model yields comparable results. 

At first glance, the F1 scores appear disappointing, especially when compared to other reported results, like \textit{Hu et Al.}, who have achieved up to $97\%$ accuracy on binary classification between normal and ROP images and $84\%$ between mild (stage 1-2) and severe (stage 3-4) images \cite{hu_et_al}. However, a few key differences render this comparison moot. First, our work and the work of \textit{Hu et Al.} use completely different datasets, and, since they have not published their dataset, there is no way to make a direct comparison between the two methods on even grounds. Second, our classifier-only experiment uses essentially the same architecture as that in \textit{Hu et Al.}'s, barring the feature aggregation step, which combines features from multiple images of the same patient to generate a single prediction. This feature aggregation operator could account for some of the difference in accuracy, but the high similarity in architecture points to the datasets as the primary factor. Lastly, their test dataset is small, consisting of only $150$ images for each of normal and ROP, and $50$ images for each of mild and severe; the result is not well-supported by sufficient data. Moreover, the methodology introduced in our work is not mutually exclusive with that of related works. We suggest that our addition of object segmentation masks be synthesized with the architectural improvements of other works to further improve the performance of ROP stage classifiers.

\section{Conclusion}
\label{sec:conclusion}

The main contribution of this paper is our addition of object segmentation to aid the CNN in classification. Towards building our hybrid system, we utilize transfer learning to take advantage of pretrained models and preprocess the images to accentuate line features, which are crucial in distinguishing between the different stages of ROP. Instead of classifying only the segmentation results or only the whole image, including the original image allows our classifier to utilize the context provided by the whole image. Through a number of careful experiments and analyses, we have demonstrated that our hybrid architecture yields a significant increase in robustness and accuracy compared to its individual components. 

To improve the overall performance, we propose a number of improvements to the dataset, including having multiple experts, instead of one, label and cross-check the images, requiring tighter bounding boxes, getting a more balanced and bigger dataset, adding normal (ROP-free) images, and incorporating multiple images of both eyes from each scan. With the improved dataset, we could modify our classifier to make predictions based on both eyes of the same patient in ways similar to the CNN pipeline used for breast cancer screening, where extracted features from each breast is concatenated to inform both breasts' prediction \cite{1903.08297}. As both eyes' ROP stages are highly correlated, this method is equally applicable here. Moreover, in light of the work of \textit{Chen et al.}, we can further improve our architecture by incorporating the fully convolutional network to generate a secondary mask, as well as replacing our single-instance classifier with a multi-instance learning module \cite{chen_zhao_zhang_wang_zhang_lei_2019} to achieve state-of-the-art performance.

\bibliographystyle{IEEEtran}
\bibliography{references}

\begin{thebibliography}{10}
\providecommand{\url}[1]{#1}
\csname url@samestyle\endcsname
\providecommand{\newblock}{\relax}
\providecommand{\bibinfo}[2]{#2}
\providecommand{\BIBentrySTDinterwordspacing}{\spaceskip=0pt\relax}
\providecommand{\BIBentryALTinterwordstretchfactor}{4}
\providecommand{\BIBentryALTinterwordspacing}{\spaceskip=\fontdimen2\font plus
\BIBentryALTinterwordstretchfactor\fontdimen3\font minus
  \fontdimen4\font\relax}
\providecommand{\BIBforeignlanguage}[2]{{%
\expandafter\ifx\csname l@#1\endcsname\relax
\typeout{** WARNING: IEEEtran.bst: No hyphenation pattern has been}%
\typeout{** loaded for the language `#1'. Using the pattern for}%
\typeout{** the default language instead.}%
\else
\language=\csname l@#1\endcsname
\fi
#2}}
\providecommand{\BIBdecl}{\relax}
\BIBdecl

\bibitem{eyewiki}
\BIBentryALTinterwordspacing
``Retinopathy of prematurity,'' Jan 2015. [Online]. Available:
  \url{https://eyewiki.aao.org/Retinopathy_of_Prematurity}
\BIBentrySTDinterwordspacing

\bibitem{tan2019deep}
Z.~Tan, S.~Simkin, C.~Lai, and S.~Dai, ``Deep learning algorithm for automated
  diagnosis of retinopathy of prematurity plus disease,'' \emph{Translational
  Vision Science \& Technology}, vol.~8, no.~6, pp. 23--23, 2019.

\bibitem{brown2018automated}
J.~M. Brown, J.~P. Campbell, A.~Beers, K.~Chang, S.~Ostmo, R.~P. Chan, J.~Dy,
  D.~Erdogmus, S.~Ioannidis, J.~Kalpathy-Cramer \emph{et~al.}, ``Automated
  diagnosis of plus disease in retinopathy of prematurity using deep
  convolutional neural networks,'' \emph{JAMA ophthalmology}, vol. 136, no.~7,
  pp. 803--810, 2018.

\bibitem{gschlieber}
A.~Gschließer, E.~Stifter, T.~Neumayer, E.~Moser, A.~Papp, N.~Pircher,
  G.~Dorner, S.~Egger, N.~Vukojevic, I.~Oberacher-Velten, and et~al.,
  ``Inter-expert and intra-expert agreement on the diagnosis and treatment of
  retinopathy of prematurity,'' \emph{American Journal of Ophthalmology}, vol.
  160, no.~3, 2015.

\bibitem{adams_2019}
G.~G.~W. Adams, ``Rop in asia,'' \emph{Eye}, Mar 2019.

\bibitem{hu_et_al}
J.~Hu, Y.~Chen, J.~Zhong, R.~Ju, and Z.~Yi, ``Automated analysis for
  retinopathy of prematurity by deep neural networks,'' \emph{IEEE Transactions
  on Medical Imaging}, vol.~38, no.~1, p. 269–279, 2019.

\bibitem{mulay_ram_sivaprakasam_vinekar_2019}
S.~Mulay, K.~Ram, M.~Sivaprakasam, and A.~Vinekar, ``Early detection of
  retinopathy of prematurity stage using deep learning approach,''
  \emph{Medical Imaging 2019: Computer-Aided Diagnosis}, 2019.

\bibitem{chen_zhao_zhang_wang_zhang_lei_2019}
G.~Chen, J.~Zhao, R.~Zhang, T.~Wang, G.~Zhang, and B.~Lei, ``Automated stage
  analysis of retinopathy of prematurity using joint segmentation and
  multi-instance learning,'' \emph{Ophthalmic Medical Image Analysis Lecture
  Notes in Computer Science}, p. 173–181, 2019.

\bibitem{dutta2019vgg}
A.~Dutta and A.~Zisserman, ``The vgg image annotator (via),'' \emph{arXiv
  preprint arXiv:1904.10699}, 2019.

\bibitem{sevik_kose_berber_erdol_2014}
U.~Sevik, C.~Köse, T.~Berber, and H.~Erdöl, ``Identification of suitable
  fundus images using automated quality assessment methods,'' \emph{Journal of
  Biomedical Optics}, vol.~19, no.~4, p. 046006, Sep 2014.

\bibitem{histogram_equalization_2019}
\BIBentryALTinterwordspacing
O.~D. Team, ``Histogram equalization,'' Dec 2019. [Online]. Available:
  \url{https://docs.opencv.org/2.4/doc/tutorials/imgproc/histograms/histogram_equalization/histogram_equalization.html}
\BIBentrySTDinterwordspacing

\bibitem{zuiderveld1994graphics}
K.~Zuiderveld and P.~S. Heckbert, ``Graphics gems iv,'' \emph{San Diego, CA,
  USA: Academic Press Professional, Inc}, pp. 474--485, 1994.

\bibitem{1703.06870}
K.~He, G.~Gkioxari, P.~Dollár, and R.~Girshick, ``Mask r-cnn,'' 2017.

\bibitem{girshick_2015}
R.~Girshick, ``Fast r-cnn,'' \emph{2015 IEEE International Conference on
  Computer Vision (ICCV)}, 2015.

\bibitem{matterport_2019}
\BIBentryALTinterwordspacing
Matterport, ``Open source implementation of mask r-cnn,'' Mar 2019. [Online].
  Available: \url{https://github.com/matterport/Mask_RCNN}
\BIBentrySTDinterwordspacing

\bibitem{szegedy2016rethinking}
C.~Szegedy, V.~Vanhoucke, S.~Ioffe, J.~Shlens, and Z.~Wojna, ``Rethinking the
  inception architecture for computer vision,'' in \emph{Proceedings of the
  IEEE conference on computer vision and pattern recognition}, 2016, pp.
  2818--2826.

\bibitem{imagenet_cvpr09}
J.~Deng, W.~Dong, R.~Socher, L.-J. Li, K.~Li, and L.~Fei-Fei, ``{ImageNet: A
  Large-Scale Hierarchical Image Database},'' in \emph{CVPR09}, 2009.

\bibitem{joan_mAP_2019}
\BIBentryALTinterwordspacing
J.~Cartucho, ``{mean Average Precision: This code evaluates the performance of
  your neural net for object recognition.}'' Sep. 2019. [Online]. Available:
  \url{https://github.com/Cartucho/mAP}
\BIBentrySTDinterwordspacing

\bibitem{1903.08297}
N.~Wu, J.~Phang, J.~Park, Y.~Shen, Z.~Huang, M.~Zorin, S.~Jastrzębski,
  T.~Févry, J.~Katsnelson, E.~Kim, S.~Wolfson, U.~Parikh, S.~Gaddam, L.~L.~Y.
  Lin, K.~Ho, J.~D. Weinstein, B.~Reig, Y.~Gao, H.~Toth, K.~Pysarenko,
  A.~Lewin, J.~Lee, K.~Airola, E.~Mema, S.~Chung, E.~Hwang, N.~Samreen, S.~G.
  Kim, L.~Heacock, L.~Moy, K.~Cho, and K.~J. Geras, ``Deep neural networks
  improve radiologists' performance in breast cancer screening,'' 2019.

\end{thebibliography}

\end{document}